**Comments on "Procedures for proper validation of record critical current density claims" by Chiara Tarantini and David C. Larbalestier, https://arxiv.org/abs/2410.22195.**

M. Polichetti[1*], A. Galluzzi[1], R. Kumar[2] and A. Goyal[2*]

[1]Laboratorio "LAMBDA" – Dipartimento di Fisica, Università di Salerno and CNR-SPIN Unità di Salerno, Fisciano (SA), Italy.

[2]Laboratory for Heteroepitaxial Growth of Functional Materials & Devices, Department of Chemical & Biological Engineering, State University of New York (SUNY) at Buffalo, Buffalo, NY, USA.

*Corresponding authors, M. Polichetti (email: mpolichetti@unisa.it), A. Goyal (email: agoyal@buffalo.edu)

*Abstract*: We provide comments on article titled "Procedures for proper validation of record critical current density claims" by Chiara Tarantini and David C. Larbalestier, https://arxiv.org/abs/2410.22195. We respond to claims and assertions that the expressions for calculation of $J_{c,mag}$ from magnetic moments proposed in Polichetti et al. 2024, https://arxiv.org/abs/2410.09197, are incorrect. We show that the analysis presented in Tarantini and Larbalestier is wrong. We also re-emphasize the expression presented in Polichetti et al. for calculation of $J_{c,mag}$ from magnetic moments in which the *pre-factor* is dimensionless and has NO units. We make the case for establishment of national user facilities for measurement of transport $J_c$ $(H,T,\theta)$ that are easily accessible in an equitable and rapid manner to enable much needed advances in the field.

Tarantini and Larbalestier [1] present and support an expression for calculation of $J_{c,mag}$ from magnetization *M* (magnetic moment per unit volume) which recalls the expression that Tallon and Talantsev [2] attributed to Gyorgy et al. [3], but they express the $\Delta M$ in $[emu/cm^3]$ whereas Gyorgy et al. explicitly state that $\Delta M$ is in $[gauss]$, and arbitrarily assign to the pre-factor 20 the dimensions of $[Acm^2/emu]$:

$$J_{c,mag}\left[\frac{A}{cm^2}\right] = 20 \frac{Acm^2}{emu}\left(\frac{\Delta M\left[\frac{emu}{cm^3}\right]}{w[cm]\left(1-\frac{w[cm]}{3l[cm]}\right)}\right) = 20 \frac{Acm^2}{emu}\left(\frac{\Delta m[emu]}{V[cm^3]w[cm]\left(1-\frac{w[cm]}{3l[cm]}\right)}\right)$$

The key message in Polichetti et al. [4] was the suggested use of the following expression for calculation of $J_{c,mag}$ from magnetization in which the number 2 has <u>NO units and is dimensionless</u>:

$$J_{c,mag} = \frac{2 \times \Delta M}{w \times \left(1-\frac{w}{3b}\right)}$$

In this formula, depending on the units in which ΔM is measured and the units of dimensions *w* and *b* in the denominator, one obtains $J_{c,mag}$ in some units which can then be converted to whatever units $J_c$ is desired in a self-consistent manner using standard tables for conversion of magnetic units. In Polichetti et al. [4], it is explicitly stated that the *pre-factor* or number 2, is dimensionless. Use of the above formula will avoid use of equations in which numbers are *assumed* to have certain units depending on the units used for the magnetic moment and for the sample dimensions, and which leads to propagation of errors. Polichetti et al. [4], suggested a formula in which the number 2 is dimensionless (*as clearly stated in the middle of page 3*) and it can be used with self-consistent magnetic units for calculation of $J_{c,mag}$ in whatever units that $J_{c,mag}$ is desired. Below we provide a detailed analysis of this summary.

The correct formula in SI is

$$J_c = \frac{2 \times \Delta\mu}{V \times w \times \left(1 - \frac{w}{3b}\right)} \quad (1)$$

If $\Delta\mu$ (where $\mu$ is the magnetic moment) is expressed in [$A\,m^2$], the volume $V = (w\,b\,d)$ is in [$m^3$], and the sample lateral dimensions $w$ and $b$ are in [$m$], the $J_c$ is expressed in SI units as $\frac{Am^2}{m^3 \times m} = \frac{A}{m^2}$. The number 2 is dimensionless.

As correctly indicated in ref. [1], it is possible to convert the magnetic moment from [$A\,m^2$] to $emu$, and the lengths from $m$ to $cm$, by means of:

$\Delta\mu[emu] = (10^3 emu/(Am^2))\Delta\mu[Am^2]$

$V[cm^3] = (10^6 cm^3/(m^3))V[m^3]$

and $w[cm] = (10^2 cm/(m))w[m]$

Using these conversions, the $J_c$ in the equation [1] can be expressed as $\frac{emu}{cm^3} \times \frac{1}{cm} = \frac{emu}{cm^4}$ with a conversion factor of

$$\frac{10^3}{10^6} \times \frac{1}{10^2} = 10^{-5}$$

with respect to the equation in SI units.

But $\frac{emu}{cm^3} = 10^3 \frac{A}{m}$ and, analogously to what has been correctly indicated in ref. [1], *if desired, the final conversion from* $\frac{A}{m}$ *to* $\frac{A}{cm}$ *is trivial*, and so we can write $\frac{emu}{cm^3} = 10^1 \frac{A}{cm}$.

In this way we obtain the units conversion from cgs emu system to some kind of "hybrid" cgs/SI system (or "practical CGS system"), as correctly identified again in ref. [1]. With this conversion, the $J_c$ in the equation [1] expressed as $\frac{emu}{cm^3} \times \frac{1}{cm}$ with a conversion factor of 10$^{-5}$, can be expressed in $\frac{A}{cm} \times \frac{1}{cm} = \frac{A}{cm^2}$ with a conversion factor of $10^{-5} \times 10^1 = 10^{-4}$, which is exactly what is indicated in the equation reported in the 3$^{rd}$ column-last row of the table 1 in ref. [1], but without the need to create a new useless "practical formula" (3$^{rd}$ column-first row) with non-coherent units that need a correction by artificially introducing a pre-factor 20 with dimensions $\frac{Acm^2}{emu}$, that does not exist in the theory.

This approach leads also to the 1$^{st}$ form of the equation 4 in ref. [4], namely:

$$J_{c,mag} = \frac{2 \times \Delta M[emu/cm^3]}{w[cm] \times \left(1 - \frac{w[cm]}{3b[cm]}\right)}$$

where it should be obvious that an expression having the dimensions of ($[emu/cm^3]/cm$) gives as result a quantity having the dimensions of $[emu/cm^4]$, definitely not practical for a current density. For this reason, the 1$^{st}$ form of the equation 4 in ref. [4] can be re-written in more practical units, by the previously indicated conversion $\frac{emu}{cm^3} = 10^1 \frac{A}{cm}$ which makes the dimensionless number 2 in the 1$^{st}$ form of the equation 4 in ref. [4] to become the dimensionless number 20 in the 2$^{nd}$ form of the equation 4 in ref. [4] and, therefore, the current density to be expressed in the more practical dimensions of [$A/cm^2$]:

$$\frac{20 \times \Delta M[A/cm]}{w[cm] \times \left(1 - \frac{w[cm]}{3b[cm]}\right)} = J_{c,mag}[A/cm^2]$$

One of the main messages of the ref. [4] is that the equation (2) there, which is the same as equation (1) above in this paper, is the correct one for the sample geometry described in the text, and it can be used independently of the fact that one has the magnetic moment in [$A\,m^2$] (and dimensions in meters) or in [emu] (and dimensions in centimetres), by using the conversion of units, without adding any non-dimensionless pre-factor:

-when the magnetic moment is in [$A\,m^2$] and the dimensions in meters, their use in equation (2) of ref. [4] leads to the equation (3) in ref. [4] where the current density is directly in [$A/m^2$]

-when the magnetic moment is in [emu] and the dimensions in centimetres, their use in equation (2) of ref. [4] leads to the 1$^{st}$ form of the equation (4) in ref. [4] where the current density is in the unpractical [$emu/cm^4$]. In order to have the current density expressed in more practical units, the conversion $\frac{emu}{cm^3} = 10^1 \frac{A}{cm}$ can be used, which introduces a multiplicative dimensionless term 10 that transform the number 2 into 20, and leads to a current density expressed in [$A/cm^2$].

The described approach is also consistent with what indicated in the textbook "Superconductivity – 2$^{nd}$ Edition, by C.P. Poole et al" [5] at page 396, where the equation to obtain the critical current density within the Bean's model are reported for several geometries. These equations are obtained by starting from the general equation (13.24) for the magnetic moment of an arbitrary shaped sample, and Poole et al. say that: "Equation (13.24) is an SI formula in which current density j is in amperes per square meter, magnetic field B is in tesla, and lengths are in meters. When practical units are used whereby j is measured in A/cm2, magnetic field in Gauss, and length in centimeters, the factor $\frac{1}{2}$ in Eq. (13.24) is replaced by 1/20. To convert the formulae below for the magnetic moment to practical units simply divide by 10."

In fact, let's start from the definition of magnetic moment $\mu$, as reported in ref. [5] -page 396, equation 13.24:

$$\mu = \frac{1}{2} \int_V [r \times j(r)] d^3r \qquad (2)$$

and for simplicity in the notation let's denote the integrated quantity as X, so: $X = \int_V [r \times j(r)] d^3r$.

The equation (2) is an SI equation, the pre-factor $\frac{1}{2}$ is <u>dimensionless</u>, and if the current density j is expressed in $A/m^2$, and the lengths in meters (m), the value of X results to be expressed in: (m) x ($A/m^2$) x ($m^3$) = (A $m^4/m^2$) = A $m^2$, coherently with the dimension of the magnetic moment $\mu$ in SI [6].

Following what reported in ref. [5], when practical units are used, the current density j is expressed in $A/cm^2$, and the lengths in centimetres (cm) the value of X results to be expressed in: (cm) x ($A/cm^2$) x ($cm^3$) = (A $cm^4/cm^2$) = A $cm^2$.

In this way, the equation (2) in practical units appears as:

$$\mu[emu] = \frac{1}{2} X[A\ cm^2] \qquad (3)$$

But: *emu*= 10⁻³ (A m²) = 10 (A cm²) [6] and so equation (3) is:

$$10\mu[A\ cm^2] = \frac{1}{2}X[A\ cm^2] \quad (3)$$

or

$$\mu[A\ cm^2] = \frac{1}{20}X[A\ cm^2] \quad (4)$$

and for the magnetization $M = \mu/V$:

$$M = \frac{\mu[A\ cm^2]}{V[cm^3]} = \frac{1}{20}\frac{\int_V [r \times j(r)]d^3r}{V}\left[\frac{A}{cm}\right] \quad (5)$$

where the current density $j$ is expressed in A/cm², and the lengths in centimetres (cm), and the pre-factor 1/20 is still <u>dimensionless</u>, coherently with the equation (4) in ref. [4].

Following this approach, as also reported by Poole et al. [5] to obtain in particular the equation 13.30 at page 396, the expressions reported in ref. [4] can be calculated with the proper units and numerical pre-factors.

Since we have analysed how to face the calculation of $J_c$ both in SI and in the 'hybrid' cgs-emu/SI system, it is worth to consider also what happens in the emu system, where electric current is not an independent physical quantity and can be expressed in abampere (*abA*), also called biot (*Bi*). More precisely [7]:

$$abA = Bi = cm^{1/2}g^{1/2}s^{-1}$$

The conversion to SI electric current unit is [7]: $1Bi = 10\ A$.

Remembering that the term "*emu*" used as unit for the magnetic moment $\mu$ is not an unit but just indicates electromagnetic units, it is important to specify that in emu system, the unit for the magnetic moment $\mu = \pi I a^2$ of a circular loop of radius $a$ carrying a current $I$ is $[erg/Gauss]$, that, of course, as indicated in the Goldfarb's paper [8] can be expressed in terms of *Bi* as:

$$[erg/Gauss] = [Bi \cdot cm^2]$$

Therefore, the equation (1) above can be used in emu system by indicating Δμ in $[Bi \cdot cm^2]$, the volume $V$ in $[cm^3]$ and the length in $w$ in $[cm]$, so leading to a current density expressed in $\frac{Bi \cdot cm^2}{cm^3} \times \frac{1}{cm} = \frac{Bi}{cm^2}$, i.e.:

$$J_{c,mag} = \frac{2 \times \Delta M[Bi/cm]}{w[cm] \times \left(1 - \frac{w[cm]}{3b[cm]}\right)} = J_{c,mag}\left(\frac{Bi}{cm^2}\right)$$

that is definitely not practical. Since $1Bi = 10\ A$, the expression above can be written also as:

$$J_{c,mag} = \frac{2 \times 10 \times \Delta M[A/cm]}{w[cm] \times \left(1 - \frac{w[cm]}{3b[cm]}\right)} = \frac{20 \times \Delta M[A/cm]}{w[cm] \times \left(1 - \frac{w[cm]}{3b[cm]}\right)} = J_{c,mag}[A/cm^2]$$

namely again the expression for $J_{c,mag}$ in practical units, were the pre-factor 20 is still <u>dimensionless</u>.

After proving the non-dimensionality of the pre-factor 20, it is probably easier to interpret the equation reported in ref. [4], and in particular the 1st and 2nd form of the equation (4):

starting from the equation (1) above in this paper, the use of the SI units system where the magnetic moment $\mu$ is expressed in $[Am^2]$ and the dimensions are in meters $[m]$ the calculation is direct and the critical current density $J_{c,mag}$ is direcly obtained in $[A/m^2]$. Since many magnetometers present their output in emu (although in several models it is possible to choose the units for the output), the same equation (1) can be used by inserting the magnetic moment $\mu$ in $emu$ and the dimensions in centimetres $[cm]$, leading for $J_{c,mag}$ a certain value (for practical reasons, let's call "Z" this value, simply resulting from the numbers given to the quantities in equation (1)) in $[emu/cm^4]$. If the value "Z" in $[emu/cm^4]$ is multiplied by 10, because of the conversion

$$\frac{emu}{cm^3} = 10 \frac{A}{cm}$$

the $J_{c,mag}$ results in $[A/cm^2]$.

This makes the 1st form of equation (4) in ref. [4] to be converted into the 2nd form of equation (4) in ref. [4], without adding any non-dimensionless pre-factors. Therefore, this means that in ref. [1] the same number 10 must be multiplied to the equation in the 4th column-last row of the Table 1, that in this way becomes identical to the equation in the 3th column-last row of the Table 1. This is also what Poole et al. [5] mean when at page 396 they write "To convert the formulae below for the magnetic moment to practical units simply divide by 10", and this is what is synthesized in ref. [4] where at page 3 it is written "In this expression, ΔM in emu/cm³ is converted to SI units with the expression 1 emu/cm³ = 10³ A/m, and so 1 emu/cm³ = 10 A/cm, yielding $J_{c,mag}$, in A/cm² ". All this just to state that if one starts from the SI equation for $J_{c,mag}$ and wants to use the emu for the magnetic moment, the only way to obtain the right $J_{c,mag}$ in $[A/cm^2]$ with the correct unit conversion is to multiply the numerical pre-factor 2 of the equation by 10.

Since the pre-factor 20 has been proven to be dimensionless, the equation (2) in ref. [1], where the pre-factor 20 has dimensions $[Acm^2/emu]$ is not correct. On the other hand, to our knowledge, no works in literature are present where these dimensions are associated to the number 20 (for the calculation of $J_{c,mag}$). And since the pre-factor 20 is dimensionless, the dimensions in equation (2) of ref. [1] for the magnetization (in $emu/cm^3$) and for the sample's lengths (in $cm$) should simply give the value of $J_{c,mag}$ in $emu/cm^4$. This has to be converted in $A/cm^2$, with the introduction of an additional multiplicative factor 10 (as shown above) that, on the contrary, is already included in the pre-factor 20. So, the pre-factor 20 in the equation (2) in ref. [1] is not compatible with the units on the left and right-hand sides in the equation (2) in ref. [1] and the attribution of the dimensions $[Acm^2/emu]$ to the pre-factor 20 to force the dimensional balance and to obtain a right value for $J_{c,mag}$ in $A/cm^2$ is not compatible with the theory (as shown above). The same confusion present in the ref. [1] about equations to use, pre-factors and dimensions, is present also in other, even very recent [2], works (although not as clearly as in ref. [1]) included those used by A. Goyal et al. [9] for the calculation of $J_{c,mag}$ from magnetic measurements, and this produced for $J_{c,mag}$ in ref. [9] a calculated value 10 times larger than the one that should have been calculated with the correct equation, pre-factor and dimensions. However, as soon as A. Goyal et al. realised the above, they requested retraction of their published paper in mid-September, which was then formally retracted on the journal website in October.

To conclude, no matter if one has the physical quantities expressed in SI, in practical units or in emu, the equation to be used (for the geometrical conditions already stated in [4]) is equation 1, and depending on the final units that are needed for the current density, one has just to use the standard conversion units [6-8], avoiding to make the error of mixing non-coherent units, and to compensate that error by artificially introducing dimensional pre-factors to force the dimensional balance of the equation. There is no need of other confusing invented equations when a well working equation, strongly supported by the theory, exists.

Otherwise it is very simple to show (we do not report this here) that for any different set of units used for the magnetic moment and for the sample's dimensions to calculate $J_{c,mag}$, a different equation, analogously to eq. (2) in ref. [1], can be created with different units (or even values) attributed to the pre-factor, in this way just producing as consequence the increase of the difficulties in comparing the results and of the possibilities to generate mistakes. We hope that this will clarify any doubt about the application of the correct equations and units for the calculation of the critical current density from magnetic measurements in the conditions detailed in ref. [4].

In order to check the validity of the discussed approach, the consistency of the measurement units and the absence of any non-dimensionless pre-factor in the expression for the calculation of the critical current density by the equation (1) above in any unit system, it can be helpful to make an example of the typical calculation of $J_{c,mag}$, if one has on hand the magnetic moment and the dimensions of the sample.

Let's consider a hypothetical sample with these characteristics:

| QUANTITY | VALUE IN EMU-CGS | VALUE IN SI SYSTEM |
|---|---|---|
| Thickness $d$ = 200 nm | $2 \cdot 10^{-5}$ (cm) | $2 \cdot 10^{-7}$ (m) |
| Width $w$ = 4 mm | $4 \cdot 10^{-1}$ (cm) | $4 \cdot 10^{-3}$ (m) |
| Length $b$ = 4 mm | $4 \cdot 10^{-1}$ (cm) | $4 \cdot 10^{-3}$ (m) |
| Volume ($wbd$) | $32 \cdot 10^{-7}$ ($cm^3$) | $32 \cdot 10^{-13}$ ($m^3$) |
| Δ(Magnetic moment) = Δ$\mu$ | 3 emu | $3 \cdot 10^{-3}$ ($A \cdot m^2$) |

Within the Bean's critical state model, for a rectangular parallelepiped sample with isotropic critical current densities and with volume $w \times b \times d$, when the magnetic field is applied perpendicular to the largest face with dimension $w$ by $b$ (with $w \leq b$) the equation to be used for $J_{c,mag}$ is the equation (1) above in the paper (or equivalently the equation (2) in ref. [4]).

So, by considering $\mu$ in $emu$ and the dimensions in $cm$, the following calculations apply:

$$J_c = \frac{2 \times \Delta\mu}{V \times w \times \left(1 - \frac{w}{3b}\right)} = \frac{2 \times 3 \, emu}{32 \cdot 10^{-7} \, (cm^3) \times 4 \cdot 10^{-1} \, (cm) \times \left(\frac{2}{3}\right)} = \frac{6 \, emu}{128 \cdot 10^{-7} \, (cm^3) \times 10^{-1} \, (cm)} \times \frac{3}{2}$$

$$= 0.0703 \cdot 10^8 \frac{emu}{cm^4}$$

Using the conversion $\frac{emu}{cm^3} = 10 \frac{A}{cm}$, this can be written as:

$$J_c = 0.0703 \cdot 10^8 \frac{emu}{cm^3} \frac{1}{cm} = 0.0703 \cdot 10^8 \cdot 10^1 \frac{A}{cm} \frac{1}{cm} = 0.0703 \cdot 10^9 \frac{A}{cm^2} = 7.03 \cdot 10^7 \frac{A}{cm^2}$$

and, since $1 \frac{A}{cm^2} = 10^4 \frac{A}{m^2}$

$$J_c = 7.03 \cdot 10^7 \frac{A}{cm^2} = 7.03 \cdot 10^7 \cdot 10^4 \frac{A}{m^2} = 7.03 \cdot 10^{11} \frac{A}{m^2}$$

On the other hand, by considering $\mu$ in ($A \cdot m^2$) and the dimensions in $m$, the following calculations apply:

$$J_c = \frac{2 \times \Delta\mu}{V \times w \times \left(1 - \frac{w}{3b}\right)} = \frac{2 \times 3 \cdot 10^{-3} \, (A \cdot m^2)}{32 \cdot 10^{-13} \, (m^3) \times 4 \cdot 10^{-3} \, (m) \times \left(\frac{2}{3}\right)} = \frac{6 \cdot 10^{-3} \, (A \cdot m^2)}{128 \cdot 10^{-13} \, (m^3) \times 10^{-3} \, (m)} \times \frac{3}{2}$$

$$= 0.0703 \cdot 10^{13} \frac{A}{m^2} = 7.03 \cdot 10^{11} \frac{A}{m^2}$$

as obtained above, and since $1\frac{A}{m^2} = 10^{-4}\frac{A}{cm^2}$

$$J_c = 7.03 \cdot 10^{11} \frac{A}{m^2} = 7.03 \cdot 10^{11} \cdot 10^{-4} \frac{A}{cm^2} = 7.03 \cdot 10^7 \frac{A}{cm^2}$$

as obtained above, without any error, incorrectness or danger in the equations, but definitely without artificially introducing any confusing dimensional pre-factors to force the dimensional balance of the equation, in this way just spreading and enhancing the confusion in literature.

The need to measure $J_{c,mag}$ from magnetic moments has arisen because of the lack of well-established and easily accessible facilities for measurements of transport $J_c$. The Department of Energy's (DOE's), Office of Electricity (OE) Program in the United States (US), supported the development of high-temperature superconducting wires and coated conductors in the US from early nineties to 2010.  The OE program also supported development of custom-built, transport $J_c$ systems which were widely available for characterization of $J_c$ $(H,T,\theta)$ to support the development of advanced methods for wire fabrication with improved vortex-pinning. In fact, essentially all previous work by one of the co-authors (Goyal) used transport-$J_c$ for all reported and seminal work on coated conductors.  However, at the end of this program most of these facilities were dismantled and there is now a dire need for national facilities for transport measurements of $J_c$ $(H,T,\theta)$ that academic users can access in an easy, equitable and rapid manner.  This is essential to enable future advances in the field in an accelerated manner.  Such national user facilities exist in other areas, for example, the DOE nanoscience user facilities (https://science.osti.gov/User-Facilities/User-Facilities-at-a-Glance/BES/Nanoscale-Science-Research-Centers), in which the majority of time on all the equipment in the facility is dedicated for *external* users, with access to the facilities decided by an *external board* and wherein all data generated belongs to the users.  The primary requirement of these national user facilities (the use of which is free and without any charges), is that the user agrees to publish the findings using the data generated in a reasonable period of time.  Academic users are especially in need of such facilities since every University effort cannot setup and maintain expensive, custom-built, transport $J_c$ $(H,T,\theta)$ systems.

Tarantini and Larbalestier postulate that the samples in Goyal et al. [9] were essentially similar to the samples in Francis et al. [10].  They state that the highest $J_c$ sample in Francis et al., SP215, was essentially similar to sample with BZO in Goyal et al. [9].  This is incorrect.  Francis et al. state that the SP215 contains 6.9vol% BZO, whereas Goyal et al., only had 2vol%BZO.  Goyal et al., also found RE substitution in BZO, which was not reported in Francis et al.  Moreover, the full composition of SP215 was not stated in Francis et al.  There are also significant differences in the chemistry and morphology of other defects in MOCVD films reported in Francis et al. and in PLD films reported in Goyal et al.  Lastly, Francis et al., do not report $J_{c,mag}$ data, only transport data.  Hence, it is not easy or straightforward to compare the results of SP215 reported in Francis et al. with that reported in Goyal et al.  It should also be noted that one of the authors (Goyal) led the CRADA between Oak Ridge National Laboratory (ORNL) and SuperPower Inc. in which the self-assembled BZO nanocolumnar defect technology was transferred to SuperPower and the first paper on this work is indicated in reference [11], leading to the commercialization of this technology.

Lastly, any superconductor film grower (for example, using pulsed laser deposition) will acknowledge that $J_c$ of films made from the same target (hence resulting in films of the same composition) with more-or-less similar epitaxial texture and microstructure, can have widely different $J_c$ 's depending on the exact deposition conditions and some deposition system dependent parameters. This additionally makes comparison of $J_c$ with SP215 difficult and inconclusive.

We agree with Tarantini and Larbalestier that ideally, characterization via transport $J_c$ $(H,T,\theta)$ is best, in particular, when geared towards power applications of superconductors. However, this requires establishment of national user facilities that can be readily accessed in a fair and equitable manner as mentioned above, especially in the US and this needs to be communicated to Program Directors at various federal agencies.

**Acknowledgements:**

R. Kumar and A. Goyal were supported by ONR Grant No. N00014-21-1-2534.

**References:**

[1] C. Tarantini and D.C. Larbalestier, "Procedures for proper validation of record critical current density claims", 2024, arXiv: https://arxiv.org/abs/2410.22195; DOI: 10.48550/arXiv.2410.22195.

[2] E. F. Talantsev and J. L. Tallon, "Fundamental Nature of the Self-Field Critical Current in Superconductors," 2024, *SSRN*. doi: 10.2139/ssrn.4978589 and https://arxiv.org/abs/2409.16758.

[3] E. M. Gyorgy; R. B. van Dover; K. A. Jackson; L. F. Schneemeyer; J. V. Waszczak, Anisotropic critical currents in $Ba_2YCu_3O_7$ analyzed using an extended Bean model, *Appl. Phys. Lett.* 55, 283–285 (1989).

[4] M. Polichetti, A. Galluzzi, R. Kumar, and A. Goyal, "Expression with self-consistent magnetic units for calculation of critical current density using the Bean's Model," 2024, *arXiv*: http://arxiv.org/abs/2410.09197. doi: 10.48550/ARXIV.2410.09197.

[5] Superconductivity – 2$^{nd}$ Edition, C.P. Poole, H.A. Farach, R.J. Creswick, R. Prozorov, ISBN 978-0-12-088761-3, Academic Press, Elsevier, 2007.

[6] https://ieeemagnetics.org/files/ieeemagnetics/2022-04/magnetic_units.pdf; https://www.nist.gov/system/files/documents/pml/electromagnetics/magnetics/magnetic_units.pdf.

[7] H.E. Knoepfel, "MAGNETIC FIELDS: A Comprehensive Theoretical Treatise for Practical Use", Wiley-Interscience Publication, John Wiley and sons inc., (2000) - ISBN-10: 0471322059; ISBN-13: 978-0471322054 – page 543.

[8] R. B. Goldfarb, "Electromagnetic Units, the Giorgi System, and the Revised International System of Units," IEEE Magnetics Letters, vol. 9, pp. 1–5, 2018, doi: 10.1109/LMAG.2018.2868654.

[9] A. Goyal, R. Kumar, H. Yuan, N. Hamada, A. Galluzzi, and M. Polichetti, "Significantly enhanced critical current density and pinning force in nanostructured, (RE)BCO-based, coated conductor," *Nat Commun*, vol. 15, no. 1, p. 6523, Aug. 2024, doi: 10.1038/s41467-024-50838-4.

[10] A. Francis, D. Abraimov, Y. Viouchkov, Y. Su, F. Kametani, and D. C. Larbalestier, "Development of general expressions for the temperature and magnetic field dependence of the critical current density in coated conductors with variable properties," *Supercond. Sci. Technol.*, vol. 33, no. 4, p. 044011, Apr. 2020, doi: 10.1088/1361-6668/ab73ee.

[11] Yimin Chen, Venkat Selvamanickam, Yifei Zhang, Yuri Zuev, Claudia Cantoni, Eliot Specht, M. Parans Paranthaman, Tolga Aytug, Amit Goyal, Dominic Lee; Enhanced flux pinning by $BaZrO_3$ and $(Gd, Y)_2O_3$ nanostructures in metal organic chemical vapor deposited GdYBCO high temperature superconductor tapes. *Appl. Phys. Lett.* 9 February 2009; 94 (6): 062513. https://doi.org/10.1063/1.3082037.